\documentclass[sigconf]{acmart}

\AtBeginDocument{%
  \providecommand\BibTeX{{%
    \normalfont B\kern-0.5em{\scshape i\kern-0.25em b}\kern-0.8em\TeX}}}

\usepackage{amsmath,amsfonts} %amssymb,
\usepackage{algorithmic}
\usepackage{graphicx}
\usepackage{textcomp}
\usepackage{xcolor}
\usepackage{multirow}
\usepackage{makecell}
\usepackage{booktabs}
\usepackage{tablefootnote}
\usepackage[normalem]{ulem}
\usepackage{comment}
\usepackage{soul}

\def\msize{0.9}

\setcopyright{acmcopyright}
\copyrightyear{2021}
\acmYear{2021}
\acmDOI{10.1145/1122445.1122456}

\acmConference[GoodIT '21]{Conference on Information Technology for Social Good}{September 9--11, 2021}{Roma, Italy}
\acmBooktitle{Conference on Information Technology for Social Good (GoodIT '21), September 9--11, 2021, Roma, Italy} 
\acmPrice{15.00}
\acmDOI{10.1145/3462203.3475918}
\acmISBN{978-1-4503-8478-0/21/09}

\begin{document}

\title{Data Collection and Labeling of Real-Time IoT-Enabled Bio-Signals in Everyday Settings for Mental Health Improvement}

\author{Ali Tazarv}
\affiliation{%
  %\institution{The Th{\o}rv{\"a}ld Group}
  \department{Electrical Eng. \& Computer Science}
  %\institution{The Th{\o}rv{\"a}ld Group}
  \institution{University of California}
  %\streetaddress{1 Th{\o}rv{\"a}ld Circle}
  \city{Irvine}
  \state{California}
  \country{USA}}
\email{atazarv@uci.edu}
 
\author{Sina Labbaf}
\affiliation{%
  \department{Computer Science}
  \institution{University of California}
  \city{Irvine}
  \state{California}
  \country{USA}}
\email{slabbaf@uci.edu}

\author{Amir M. Rahmani}
\affiliation{%
  \department{School of Nursing, Computer Science}
  \institution{University of California}
  \city{Irvine}
  \state{California}
  \country{USA}}
\email{a.rahmani@uci.edu}

\author{Nikil Dutt}
\affiliation{%
  \department{Computer Science}
  \institution{University of California}
  \city{Irvine}
  \state{California}
  \country{USA}}
\email{dutt@uci.edu}

\author{Marco Levorato}
\affiliation{%
  \department{Computer Sciecne}
  \institution{University of California}
  \city{Irvine}
  \state{California}
  \country{USA}}
\email{levorato@uci.edu}

\renewcommand{\shortauthors}{Tazarv, et al.}

\begin{abstract}
Real-time physiological data collection and analysis play a central role in modern well-being applications. Personalized classifiers and detectors have been shown to outperform general classifiers in many contexts. However, building effective personalized classifiers in everyday settings - as opposed to controlled settings - necessitates the online collection of a labeled dataset by interacting with the user. This need leads to several challenges, ranging from building an effective system for the collection of the signals and labels, to developing strategies to interact with the user and building a dataset that represents the many user contexts that occur in daily life. 
Based on a stress detection use case, this paper (1) builds a system for the real-time collection and analysis of photoplethysmogram, acceleration, gyroscope, and gravity data from a wearable sensor, as well as self-reported stress labels based on Ecological Momentary Assessment (EMA), and (2) collects and analyzes a dataset to extract statistics of users' response to queries and the quality of the collected signals as a function of the context, here defined as the user's activity and the time of the day.
%The labeling of IoT-generated real-time data in everyday-settings in a long-time periods (a few months) poses significant challenges due to user behaviour effects and limited signal quality. 
%In this study, we introduce a layered system architecture that supports a tunable collection of data samples for labeling and presents a method for selecting informative samples from the stream of real-time data for labeling.
%We sent out questionnaires and collected responses (i.e. instant stress levels, physical activity, etc.) from fourteen volunteers over a period of time (between 1 week to 3 months) to test our labeling query engine.  
%We present results on interactions with subjects in everyday settings in long-time periods, including the interactions with the participants, user behaviour, response times and response rates in different contexts.
%We evaluate the quality of data and temporal correlations of the signals (in cases where subjects answer with delay) in various contexts.
%Other than that we analyze how the portion of data that is labeled reasonably represents the entire dataset.
%Results show how promising the collected labeled dataset  can be in everyday settings, the portion of data that is labeled is capable of representing the entire dataset, and quality of the samples and the user response-delay are a function of the activity they users are involved in (context).
%Our study lays the groundwork for more sophisticated labeling strategies that generate context-aware, personalized models that will empower health professionals to provide personalized interventions.
\end{abstract}




\maketitle

\section{Introduction}
\label{sec: intro}
%The term "stress" as it is currently used was defined by Hans Selye as "the non-specific response of body to any demand for change" \cite{selye-hans}. Aside from the natural environment in which our stress reaction system was originally evolved in it, in the new industrial world people are experiencing increased levels of constant stress coming from different sources including workplace \cite{bakker2011s}.\\ % ,paoli2001third
This study stems from the UNITE project housed at the University of California, Irvine. The goal of the project is to improve the well-being of pregnant women in underrepresented communities by integrating in-home visitations with wearable-based fine grain monitoring and interventions. The focus is on
detecting and mitigating stress, a key indicator 
%predictor 
of pregnancy outcomes.
%Maternal Care (MC) aims to improve the overall health and welfare of an unborn child, pregnant woman, and the entire family \cite{kikuchi2015effective}.
%The quality of MC in the US depends on individual's socioeconomic status (SES), access to healthcare, end their education \cite{cokkinides2001health}; consequently underserved communities with low SES do not receive satisfactory MC, resulting in poor birth outcomes and a decline in mother's and family's overall health \cite{meyer2016working}.
%A unique integrated maternal home visitation and group health education approach developed by a non-profit agency has provided prenatal and postnatal services to pregnant women and families in underserved communities and has demonstrated improved MC outcomes \cite{guo2016community}.
%However, many gaps exist in the current model, presenting an opportunity to use technology to improve the quality of MC through ubiquitous monitoring, and expand the community beyond self-selection.

Within this context, in this paper we analyze the feasibility of an online learning strategy, whose final objective is building a stress detector by collecting a labeled dataset associating physiological signals to stress labels. Different from prior studies, we focus on every day settings, where there are no restrictions on movements and the environment, and subjects are doing normal daily routines. 

In addition to the development of an effective system for the real-time collection of data, this scenario presents several inherent challenges related to the data toward the training of effective classifiers. Intuitively, one of the key challenges is the quality of the physiological signals, which may depend on several factors, including motion, and thus activity~\cite{han2020objective,NAEINI2019551}. However, a critical aspect is the response of the user to the queries (EMA). The user context may influence willingness to respond and the response time, thus affecting how representative the dataset is of the user's activities and the correlation between samples and labels.

This paper makes the following contributions:

\noindent
$\bullet$ We build and deploy a three-tiered system for the collection and real-time analysis of labeled stress data. The system is composed of wearable sensors, an edge layer and a cloud server.
%Sensors are located on a wearable device (i.e. smart watch) which is worn by subjects on their wrist. Sensors collect biosignals (PPG) and movement data, and send those to the cloud either directly through Wi-Fi (if available) or through the internet on the cell phone. A Smart phone is necessary to display the surveys and ask for the labels. The smart phone also connects to the smart watch and can receive the data and transfer it to the cloud (in case the watch is not directly connected to Wi-Fi). The cloud server receives the raw signals from sensors, analyses and determines for what samples it is necessary to ask for labels.
We discuss system-level challenges that impact data acquisition capabilities.

\noindent
$\bullet$ While the tiered system is capable of acquiring a large number of physiological signal samples, the user may be willing to label only a small fraction of them. We then develop a strategy to request labels based on the signal itself even in the absence of a prebuilt classifier. The strategy aims to collect a number of samples proportional to the density of samples in the feature space, but also capture outliers and rare events.

\noindent
$\bullet$ We collect a dataset from a group of volunteers in everyday settings. Specifically, we collect biosignals photoplethysmogram (PPG) along with the movement data -- Acceleration, Gyroscope and Gravity -- from sensors on a smart watch in a real-time scheme, and collect self reported stress levels from participants. 
The raw signals (PPG, ACC, Gyro and Gra) are collected in a window of 2 minutes, once every 15 minutes. 
The list of labels we collect includes mental stress level, emotional status and physical activity.
%We collect the stress ranking labels in 5 levels: \textit{not at all}, \textit{a little bit}, \textit{some}, \textit{a lot}, \textit{extremely}. Data collection took place during Summer and Fall 2020. User acceptability of interactive solutions results in a trade-off between the number of collected labels from the subjects vs. the accuracy of the predictive model. This trade-off raises the importance of a smart label query strategy.

\noindent
$\bullet$ We study the quality of the signals and the willingness of the user to respond to queries as a function of the current context, here defined in terms of user's activity and the time of the day. We also show distribution plots of response time and response rate of users, during our experiment. 
On average the response rate is usually lower in early morning (7-9 AM) and higher in early afternoon (2-3 PM), and response time shows a different distribution for different activities.
We analyze the quality of PPG signals for each predefined activity separately, and show that during the activities with less movements, the measured signal has better quality.
We analyze the temporal correlation of samples as a function of context (activities) and show that consecutive samples are more similar to each other, compared to distant samples.

The rest of the paper is organized as follows:
%In order to collect the data and send queries for labels optimally,
This paper starts by describing the system model and architecture used for data collection and sending queries for labels, highlighting the challenges introduced by the system constraints in Section \ref{sec: system_model}. 
Section \ref{sec: data_collection} describes our strategy for sending queries and collecting labels for the data optimally in the absence of a prebuilt classifier.
%It also explains the methods and metrics we used for signal analysis.
Section \ref{sec: methods} describes our approach and the methods we used for analysis of the collected data. 
Section \ref{sec: data_analysis} presents the results of the analysis, in terms of coverage of sample space with labeled data, contextual analysis, signal quality, and user behaviour analysis in different contexts and on different self reported stress levels.
Finally, Section \ref{sec: conclusion} concludes the paper with a summary and directions for future~work.
\vspace{-1em}
\section{System Model}
\label{sec: system_model}
\begin{figure}[tbp]
    \centering
    \includegraphics[width= 1 \linewidth]{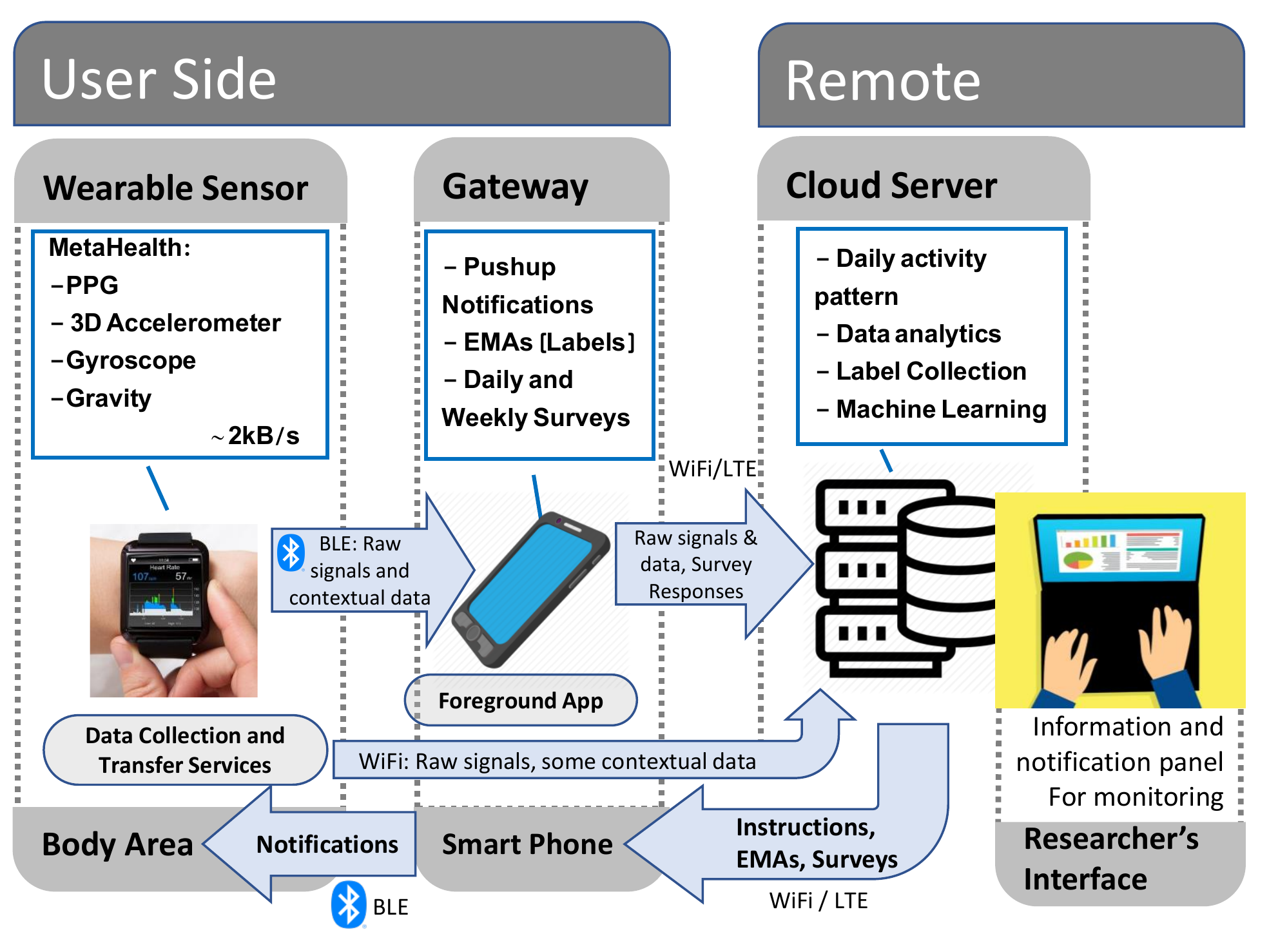}
    \vspace{-2em}
    \caption{Overview of the system architecture.}
    \vspace{-1em}
    \label{fig: architecture}
\end{figure}

First, we describe the system we developed to enable data collection and real-time interaction with the user. As shown in Fig.~\ref{fig: architecture}, the system is composed of three tiers: sensor layer, edge layer and cloud layer. The sensor layer which is a smart watch, collects the raw signals while the cloud layer performs feature extraction and other computationally expensive and power consuming tasks, including selecting a portion of the samples to be labeled by the user.

We provide users with an interface to report labels in the form of a smart-phone application. 
The smartphone app asks the participants to label the samples through an Ecological Momentary Assessment (EMA), in which push notifications queries the participant about their stress level, recent physical activity or physical state (e.g. sitting, standing, etc.).
The phone also functions as a gateway, building a connection path for the smart watch to transfer the sensor data to the cloud layer (through the internet connection on the phone).
In the following, we describe each layer in detail and discuss the challenges we encountered for data collection using wearables in everyday settings.

\subsection{Sensor Layer}
%In order to collect data, process, and provide EMAs to participants we needed to develop a real-time platform to complete this control loop. The biggest challenges were that most available platforms do not provide us with raw PPG signal or they are not giving us realtime data processing options. So, we needed to find a platform that can face this two challenges. This section will be formed in three main stages of process: a. data collection, b. data processing, c. EMA
We selected a wearable platform capable of acquiring and transmitting raw physiological (PPG) and motion (Accelerometer, Gyroscope and Gravity) signals. Specifically, we use Samsung Gear Sport smartwatches and developed a service in Tizen that is capable of collecting raw PPG, accelerometer, gyroscope and gravity signals and sending them to the cloud in real-time. The sampling frequency of the sensors is $20 Hz$. The watch can send the data directly to the cloud layer if it is connected to a local Wi-Fi (path 1) or through through the internet connection on the smartphone (path I) as explained later.

The data collection application on the sensor layer includes two services and a user interface (UI). The first service collects the sensor data at a constant rate (once every 15 minutes) and duration (2-minute intervals) and sends it to the cloud in real-time. If the service fails to send the data immediately, the data sample is stored on the watch and transferred to the server at a later time. The UI is a simple app on the watch for restarting these two services. 

\subsection{Cloud Layer}
A cloud web-server receives the data samples from the watch and immediately initiates processing.
Based on the observed features of each incoming sample, an internal logic determines whether or not to trigger the EMA to collect a label from the user. If triggered, it sends a signal to the smartphone and a push notification appears on the screen, if the user opens it, the UI (a phone app) shows a questionnaire composed of simple questions corresponding to labels. The responses are then transferred to the cloud. The samples, features and labels are stored in a MongoDB database.

\subsection{Edge Layer}
The sensor is connected to a smartphone via Bluetooth Low Energy (BLE). If the watch is not connected to a local Wi-Fi, in order to send the collected data to the cloud, the watch proxies the phone's internet connection (path 2) through BLE. This setting is energy efficient, and thus suitable for everyday setting applications. This back up connection route is designed to take effect when the watch is not directly connected to a local Wi-Fi router. Additionally, we designed a UI for the smartphone (android and iOS apps) in order to communicate with the users and collect the labels.

\subsection{Challenges}
From a system perspective, the first design challenge is to set the monitoring duration and frequency such that: 1.~The total delay from data acquisition to the EMA notification is tolerable, 2.~Power consumption matches the characteristics of the devices and user requirements, 3.~The sample quality allows meaningful feature extraction, and 4.~The signals provide a meaningful representation of temporal variations. Note that the capacity of the wireless channel connecting the watch to the smartphone is extremely limited.
\subsubsection{End-to-End Delay}
First, we measured the bandwidth of the BLE proxy Internet connection manually to characterize hardware limitations. The maximum data-rate is approximately $80 kB/s$. As the sensor generates $2 KB/s$ of data, the upload time for each minute of data is only $1.5$s. The processing time at the server is negligible compared to the upload time.

After collecting and sending the EMA, there is a small delay until the notification appears on the phone. This delay depends on internet connection, type of the phone and the mode the phone is operating on (power saving, etc.). However, this delay is typically small compared to the users' response time.

\subsubsection{Power Limitations on the Watch}
To measure power consumption, we first used the watch without any monitoring services and measured the battery lifetime, and compared it with that observed with the monitoring system  continuously active. The measured battery lifetimes are $\sim40$ hours and $\sim10$ hours, respectively. In order to extend battery lifetime to $24$ hours (to allow for nightime recharging), we then need to keep the monitoring system active for at most $22\%$ of the time. This limits our ability to collect continuous signals over extended periods of time, and raises the issue of how to shape a parsimonious sample collection strategy. 

\subsubsection{Sampling Times and Signals Duration}
We determine sample collection based on the considerations above. Importantly, the data collection window plays an important role in the quality of PPG signals. However, if the window duration is increased, then we are forced to do the measurement fewer times throughout the day. As shown in \cite{Baek2015}, a 2-minute time window of PPG/ECG signal can provide us with sufficiently accurate extraction of the majority of Heart Rate Variability (HRV)\footnote{Heart Rate Variability or HRV is a set of informative features including statistics of heart beats that can be extracted from PPG signals.} features. Hence, By setting the minimum duration of data samples to 2 minutes, sampling every 15 minutes will satisfy all system constraints described above while extending the battery life to up to 42 hours in practice.

\section{Data Collection}
\label{sec: data_collection}
The ultimate goal of data collection is to train personalized classifiers that can detect mental health status based on biosignals (PPG) and signals describing motion (Accelerometer, Gyroscope and Gravity).
%As mentioned before, the watch measures 2 minutes of data out of every 15-minute interval, and sends it immediately to the cloud (if possible, either through first service or through the phone). 
One of the key challenges in collecting such datasets in everyday settings is the interaction with the users, as sending queries for labeling too often can be overwhelming and may lower response rate and eventually degrade the dataset. Intuitively, the system should parsimoniously trigger the EMA to collect a label to build a meaningful dataset as quickly as possible without imposing excessive burden on the user.
To this aim, we devised a selection method that triggers the labeling query based on statistics of the previously collected and current samples.

\begin{figure}
    \centering
    \vspace{-1em}
    \includegraphics[width= 1 \linewidth]{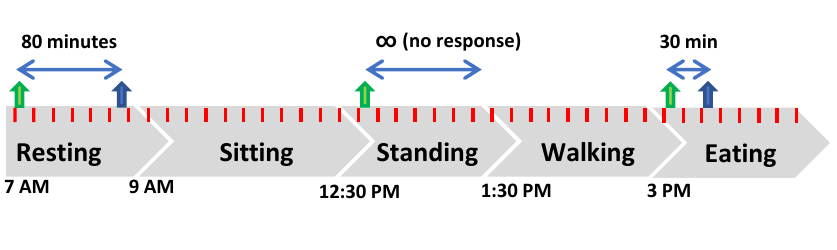}
    \vspace{-3em}
    \caption{Example of Data collection in everyday setting.}
    \vspace{-1em}
    \label{fig: my_label}
\end{figure}

Figure \ref{fig: my_label} represents such a data collection scenario. Samples are collected once every fifteen minutes (red pointers), green arrows correspond to the EMA notifications, and the vertical blue arrows correspond to responses from the user.

The list of labels we collect through the EMA includes stress levels (\textit{not at all}, \textit{a little bit}, \textit{some}, \textit{a lot}, and \textit{extremely}), emotions (\textit{sad, mad, neutral or happy}), and recent activities or physical status (e.g., sitting, walking, jogging, etc.).
Later in this section we describe how we select a portion of samples to be labeled.

\subsection{Data Cleaning and Feature Extraction}
Before we apply the selection method, we pre-process the raw signals and extract the corresponding features (using HeartPy Package \cite{van2018heart}).
When the raw PPG sample is received at the cloud layer, it is first passed through a Butterworth band-pass filter to clean up the high and low frequency noises. The band-pass filter is of order 3, with cut off frequencies set at $(0.7Hz$, $3.5Hz)$, corresponding to $42bpm$ and  $210bpm)$ respectively. Then the signal is passed through a moving average filter (length = 0.75 seconds). We, then, apply a peak detector to the filtered signal. Using the peak points of the filtered signal, we extract thirteen features from each sample (window). These features are:
BPM, IBI, SDNN, SDSD, RMSSD, pNN20, pNN50,
MAD,
SD1, SD2, S, SD1/SD2, and BR\footnote{
\textbf{BPM:} Beats per Minute, Heart Rate.
\textbf{IBI:} Inter-Beat Interval, average time interval between two successive heart beats (called NN intervals).
\textbf{SDNN:} Standard Deviation of NN intervals.
\textbf{SDSD:} Standard Deviation of Successive Differences between adjacent NNs.
\textbf{RMSD:} Root Mean Square of Successive Differences between the adjacent NNs.
\textbf{pNN20:} The proportion of successive NNs greater than 20ms (or 50ms for pNN50).
\textbf{MAD:} Median Absolute Deviation of NN intervals.
\textbf{SD1 and SD2:} Standard Deviations of the corresponding Poincaré plot.
\textbf{S:} Area of ellipse described by SD1 and SD2.
\textbf{BR:} Breathing Rate.}.
We use these features for further processing and decision makings.

\subsection{Strategy for Labeling Selected Data}
Data collection consists of an Initial Phase and a Query Phase:

\textbf{Initial Phase:} 
We start the procedure by observing the first N samples for each subject to get an estimate of the distribution of samples in the sample space for that subject. In this phase we do not collect any labels. We used N = 100 in our experiment.
It is important to note that different subjects might have different patterns in their PPG which can result in personalized distributions in the sample space. This motivates the need to estimate the distribution of the data for each subject separately.

\textbf{Query Phase:}
For samples after the initial phase, we trigger the EMA for a subset of samples. 
The probability of triggering the EMA for each sample is proportional to the density of the region of the sample. This way, if a sample falls in a region in which there has been a large number of unlabeled samples it is more likely that we trigger the EMA. For each region, after we collect sufficient number of labels, we stop collecting labels. However, all the probability values (for all the regions) are \textit{clipped} on the bottom at P = 0.1. So if a samples falls in a region where there is little or no previous samples, the probability of query is still non-zero. This results in exploring unseen regions, as well as more dense regions.
%
% \noindent
% To obtain the partitioning boundaries, for each feature we set them as $\mu\pm\frac{\sigma}{2}$, in which $\mu$ is the mean value and $\sigma$ is the standard deviation of each feature among the collected samples. After every 100 recorded samples ($\sim$25 hours), the distribution boundaries and the distribution probability values are updated (fine adjustments).

The ultimate goal of this experiment is to train a personalized classifier to predict mental health conditions. In order to do that, we need to collect enough labeled data, such that they optimally cover the sample space.
%One trivial idea is to pick a portion of these samples randomly and uniformly, query for their labels. One problem with this approach is that the denser a region (bin) in the sample space is, the higher the number of labeled samples will be from those bins. On the other hand, regions with lower densities, will have lower (if any) number of labels. Therefore, the majority of labels will be from denser bins, and some bins might remain with just a few labels, or even empty, both of which might result in high prediction errors. On the other hand, a large number of labels from high density regions, is probably unnecessary after a certain point. So this way, we are not using the labeling budget efficiently.
%In our method, for regions with low densities, we assign a minimum probability ($p_0 = 0.05$), which helps collect labels for all the bins with low (or zero) probability too. 
To measure how the labeled samples cover the sample space we need a measure.

%\marco{maybe these definitions could go in the section before to better define the sampling strategy? A technical description would be best} 
If we define all the incoming samples as $X = \{x_0, x_1, ..., x_M\}$, labeled samples as $U_i = \{u_0, u_1, ..., u_i\}$ ($U_i \subset X$) and the corresponding labels as $Y_i = \{y_0, y_1, ..., y_i\}$ ($Y_i$ is the set of labels collected up to label number \textit{i}), we then define the coverage metric as:
\[
F_D(i) = \frac{\mathbf{card(}\,\{x \in X\ |\,\, \|x - u^*\|>D\}\,\mathbf{)}}{\mathbf{card(X)}}
\]
in which $u^*$ is the closest labeled sample to the sample $x$, \textbf{card} is the cardinality of the set, and $D$ is a distance constant. Note that $u_k$ or $x_k$ are each a vector of 13 elements (features), extracted from a PPG sample (window size 2 minutes).
As we collect more labels, we count the number of samples which are farther than a certain threshold $D$ to the closest labeled sample collected up to that point. The ratio between this number and the total number of samples gives us a metric $F_D(i) \in [0,1]$. The smaller this metric, the better labeled samples represent the entire data.

\section{processing and evaluation Methods}
\label{sec: methods}
%Signal quality (amount of motion and noise artifacts) is an important factor to consider when collecting data in an everyday setting.  
\subsection{Signal Quality}
\label{subsec: methods_quality}
The quality of PPG signal is among the most important factors to consider when collecting data in everyday settings. Due to the architecture of PPG sensors, PPG signals are highly prone to motion and noise artifacts (MNA) \cite{maeda2011relationship}, which can make them unreliable, especially in everyday-settings. Therefore, a signal quality assessment on the collected signal is essential.

Several quality assessment indices for PPG signals are proposed in the literature. In this study we use five different indices \cite{mahmoudzadeh2021lightweight}: Variation in Skewness of Heart Cycles measures the variations in asymmetry in distribution of peaks in the heart cycles; Variation in Kurtosis of Heart Cycles evaluates the variations in flatness or peakedness level of the heart cycles; Variation in approximate Entropy of Heart Cycles evaluates the variations in complexity of the heart cycles; Shannon Entropy obtains the level of noise in a segment of PPG signal; and Spectral Entropy calculates the signal complexity in frequency domain. For all of these five indices the lower values indicate a more \textit{reliable} signal. These indices are extensively described in \cite{mahmoudzadeh2021lightweight} with formal definition for each. 

\subsection{Activity Detection}
We collect data from Accelerometer, Gyroscope, and Gravity sensors as well as the PPG signal. Each of these signals (except PPG) are measured in 3 dimensions $(x,y,z)$ and can be used to detect user's activity during each measurement.
The MNA in PPG signals show different patterns depending on users' activities. So if we partition samples based on the type of activity, data in each partition might be easier to analyze.

To do that, we need an activity detector. We used a publicly available dataset \cite{weiss2019wisdm} and trained a Random Forest classifier on this dataset to build an activity detector. The external dataset consists of accelerometer and gyro data from 51 subjects, and labeled with 18 different activities. We selected the activities that are expected to be more common in everyday settings (\textit{sitting, standing, walking, and Jogging}), kept these labels and changed the rest of them to \textit{others}. We then trained our activity detector on this dataset. After fine tuning the model parameters, the activity detector showed $84\%$ accuracy on leave-two-subjects-out evaluation method (trained on 49 subjects and tested on two subjects). We then used this trained classifier to predict the dominant activity during the measurement of each sample in our dataset. Then we partition our collected data into subsections based on those activities. The distribution of predicted activities on the entire dataset is shown in Table \ref{tab: partitioning}.

\begin{table}[bt]
    \centering
    \vspace{.15cm}
    \caption{Size of partitions based on predicted activity}
    \vspace{-1em}
    \begin{tabular}{l r r r r r r} 
    \hline \\[-1.5ex]
    ACTIVITY: & Sit & Stand & Walk & Jog & Others & Total \\
    %\rotatebox{-90}{sitting} & \rotatebox{-90}{standing} & \rotatebox{-90}{walking} & \rotatebox{-90}{jogging} & \rotatebox{-90}{others} & \rotatebox{-90}{total}\\
    \hline \\[-1ex]
    Samples: & {\small31,040} & {\small 5,919} & {\small3,849} & {\small12} & {\small33,254} & {\small 74,074}\\
    Percent: & {\small\%41.9} & {\small\%8.0}& {\small\%5.2}& {\small \%0.016} & {\small\%44.9} & {\small\%100}
    \end{tabular}
   \label{tab: partitioning}
   \vspace{-2em}
\end{table}

\section{Data analysis and Results}
\label{sec: data_analysis}
We proposed a labeling query engine that determines (in real-time) what samples are interesting to be labeled by the user. 
We collect raw biosignals (PPG) and raw movement data in durations of 2 minutes, up to 4 times per hour from fourteen volunteers (ten males and four females) over periods of time between 1 week to 3 months for different subjects. The corresponding labels come from self reported EMAs which are triggered for some samples.
Table \ref{tab: data_count} presents the total number of samples we have collected throughout the experiment, along with the number of labels, and the number of labels that could be assigned to a sample (labels that had a sample withing 16 minutes around them).
%S1: 148, S2: 160, S3: 197, S4: 206, S5: 265, S6: 446, S7: 470, S8: 533,
%S9: 538, S10: 552, S11: 656, S12: 729, S13: 749, S14: 761
\begin{table*}[bt]
    \centering
    \vspace{-1em}
    \caption{Number of samples and labels for each subject}
    \vspace{-.5em}
    \begin{tabular}{l r r r r r r r r r r r r r r r} 
    \hline \\[-1.5ex]
    \textbf{Subject} & \textbf{S01}   & \textbf{S02}   & \textbf{S03}   & \textbf{S04}   & \textbf{S05}   & \textbf{S06}   & \textbf{S07}   & \textbf{S08}   & \textbf{S09}   & \textbf{S10}  & \textbf{S11}  & \textbf{S12}  & \textbf{S13}  & \textbf{S14} & \textbf{Total} \\
    \hline\\[-1.5ex]
\textbf{Samples} & 4,580 & 2,164 & 1,764 & 2,580 & 2,267 & 17,552 & 10,087 & 2,752 & 1,236 & 7,910 & 2,555 & 12,296 & 3,738 & 1,332 & 74,074 \\
\textbf{Total labels} & 228 & 101 & 67 & 56 & 68 & 376 & 105 & 96 & 53 & 119 & 73 & 956 & 47 & 61 & 2,406\\
\textbf{Used labels} & 217 & 92 & 42 & 53 & 59 & 370 & 101 & 93 & 50 & 104 & 60 & 942 & 45 & 55 & 2,283 \\[1ex]
    \hline \\[-1.7ex]
    \end{tabular}
   \label{tab: data_count}
\end{table*}

We now analyze the dataset, using the methods and metrics defined in Section \ref{sec: methods}. Analysis include coverage of sample space, temporal correlations of samples, quality of PPG signals, 
and response time/rate of subjects in different contexts in everyday settings.

\subsection{Sample space Coverage}
\begin{figure}[tbp]
    \centering
    \vspace{-1.5em}
    \includegraphics[width= \msize \linewidth]{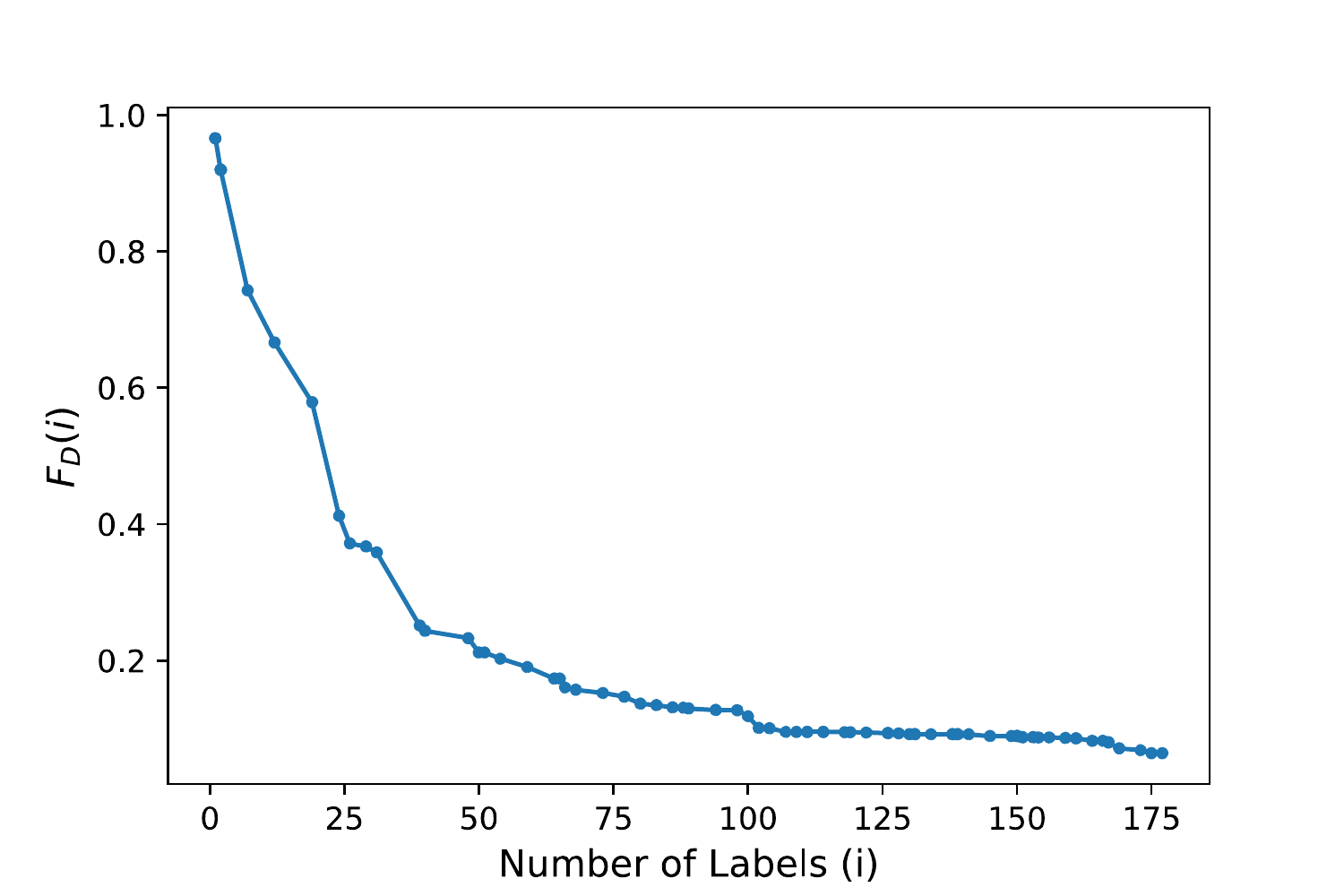}
    \vspace{-1em}
    \caption{Fraction of unlabeled data that are farther than distance D = 1.5 from the closest labeled point.}
    \vspace{-1em}
    \label{fig: metric}
\end{figure}

%The ultimate goal of data collection is to train a classifier based on these signals to perform predictions on human health and well-being. In order to get an accurate classifier, we need to collect enough labeled data, such that they optimally and proportionally cover the sample space. 
In section \ref{sec: data_collection} we defined a coverage metric $F_D(i)$ that quantifies how well the labeled samples cover the sample space. Labeled samples that cover the sample space proportionally and optimally are an important requirement for training a classifier that predicts human health and well-being.
%\marco{Any explanation or reference to motivate this statement?}
A plot of this metric over the number of labeled samples for one subject and for the threshold distance $D=1.5$ is presented in Figure \ref{fig: metric}. In this plot the sample space coverage metric goes from around 1 to less than 0.1 as we collect the first 100 labels (generally it depends on the response rate but for this user it took about 3 weeks).

\subsection{Temporal correlation of samples}
In the procedure of collecting labels, sometimes users cannot respond to the push notifications quickly enough for the sample to be meaningful.
One important question then is how quickly humans' status (e.g., stress level, emotional status, and the corresponding physiological effects) changes. We submit queries for labels only a few times each day and obtain responses for a fraction of them. As a result, the correlation between the sample and the label may degrade. Thus, we perform a temporal analysis of the collected samples for several contexts. 
%We try to observe the distribution of distances between samples in the sample space, for samples that are measured at certain time lengths apart from each other.
Specifically, we observe how the average distance of samples evolves in the sample space over time. 

\begin{figure}[tbp]
    \centering
    \vspace{-1.5em}
    \includegraphics[width= \msize \linewidth]{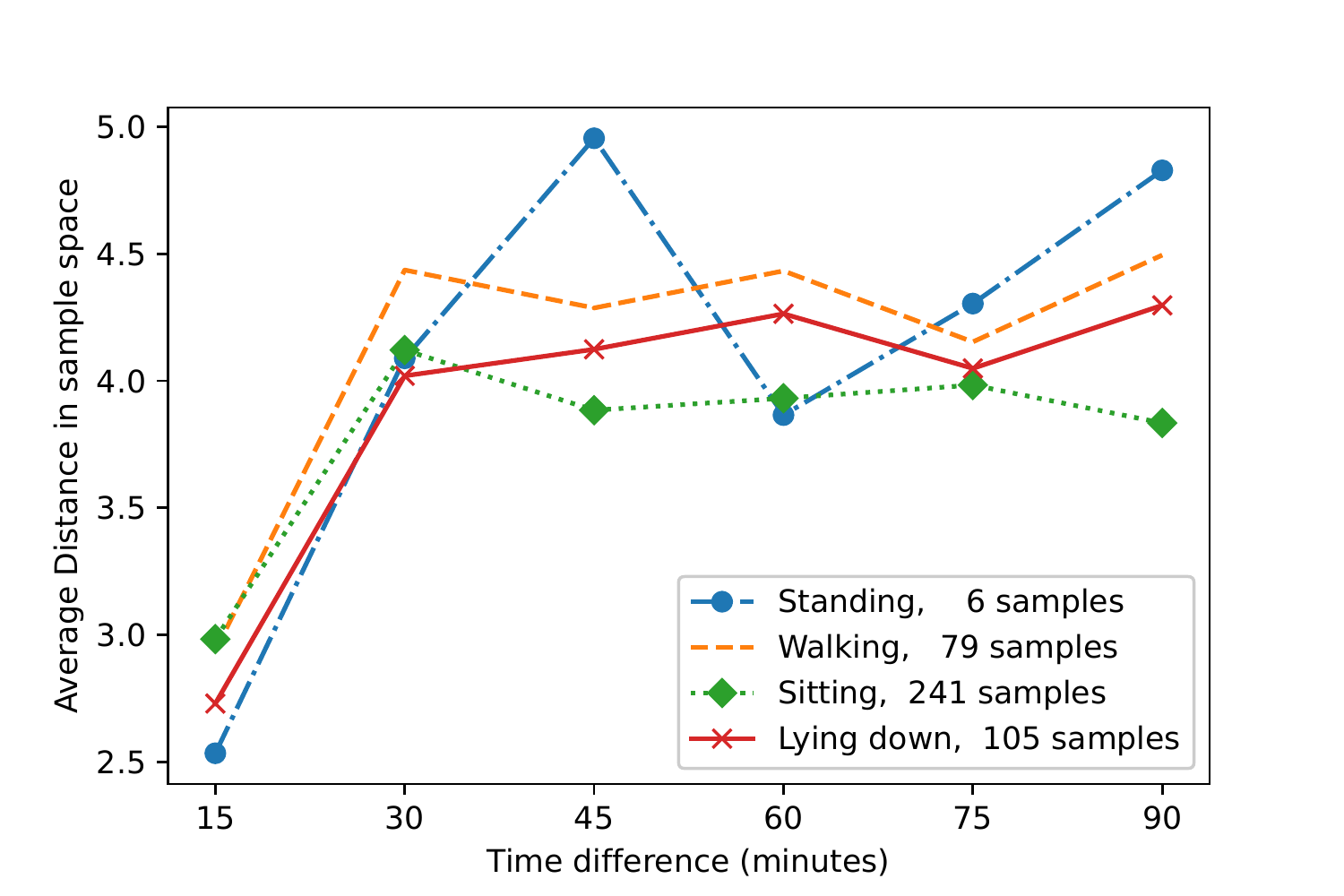}
    \vspace{-1em}
    \caption{Average distance of samples (after normalizing the features) VS. their time distance for subject S12.}
    \vspace{-1.5em}
    \label{fig: temporal_activity}
\end{figure}

\begin{figure}[tbp]
    \centering
    \includegraphics[width= \msize \linewidth]{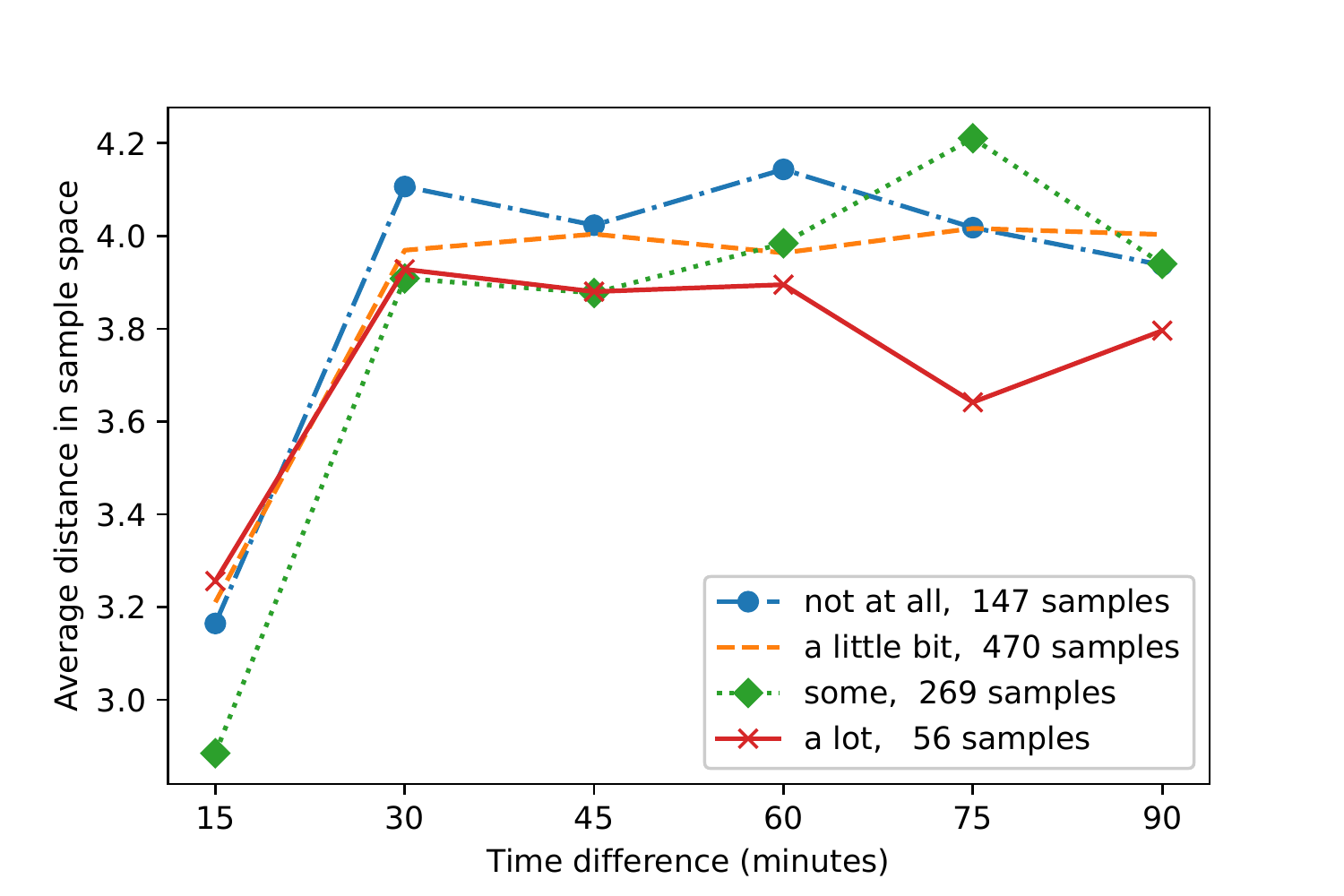}
    \vspace{-1em}
    \caption{Average distance of samples VS. their time difference, for subject S12, for different reported stress levels.}
    \vspace{-1.5em}
    \label{fig: temporal_stress}
\end{figure}

Two plots of the average distance of samples that are T minutes apart are presented in Figures \ref{fig: temporal_activity} and \ref{fig: temporal_stress} for various contexts and stress levels, on the data from one subject.
The important takeaway from these two plots is that consecutive samples (15 minutes apart) are similar to one another, but for samples that are farther than 15 minutes apart, there is no significant or consistent similarity (the average distance almost saturates at 30 minutes and after). This pattern is consistent among various contexts, various stress levels, and also various subjects.

In summary, we observe that different activities and stress levels influence the coherence of the collected samples, which may affect the ability of the system to detect stress levels in various activities and user situations.

\subsection{Data Quality Analysis}
Quality of PPG signals is one of the biggest challenges whilst collecting data in everyday-settings, therefore a quality analysis of the collected data is essential. We use the five SQIs introduced in section \ref{subsec: methods_quality} for the quality assessment of samples, and perform a separate analysis for each activity.

\begin{figure}[tbp]
    \centering
    \vspace{-1.5em}
    \includegraphics[width= \msize \linewidth]{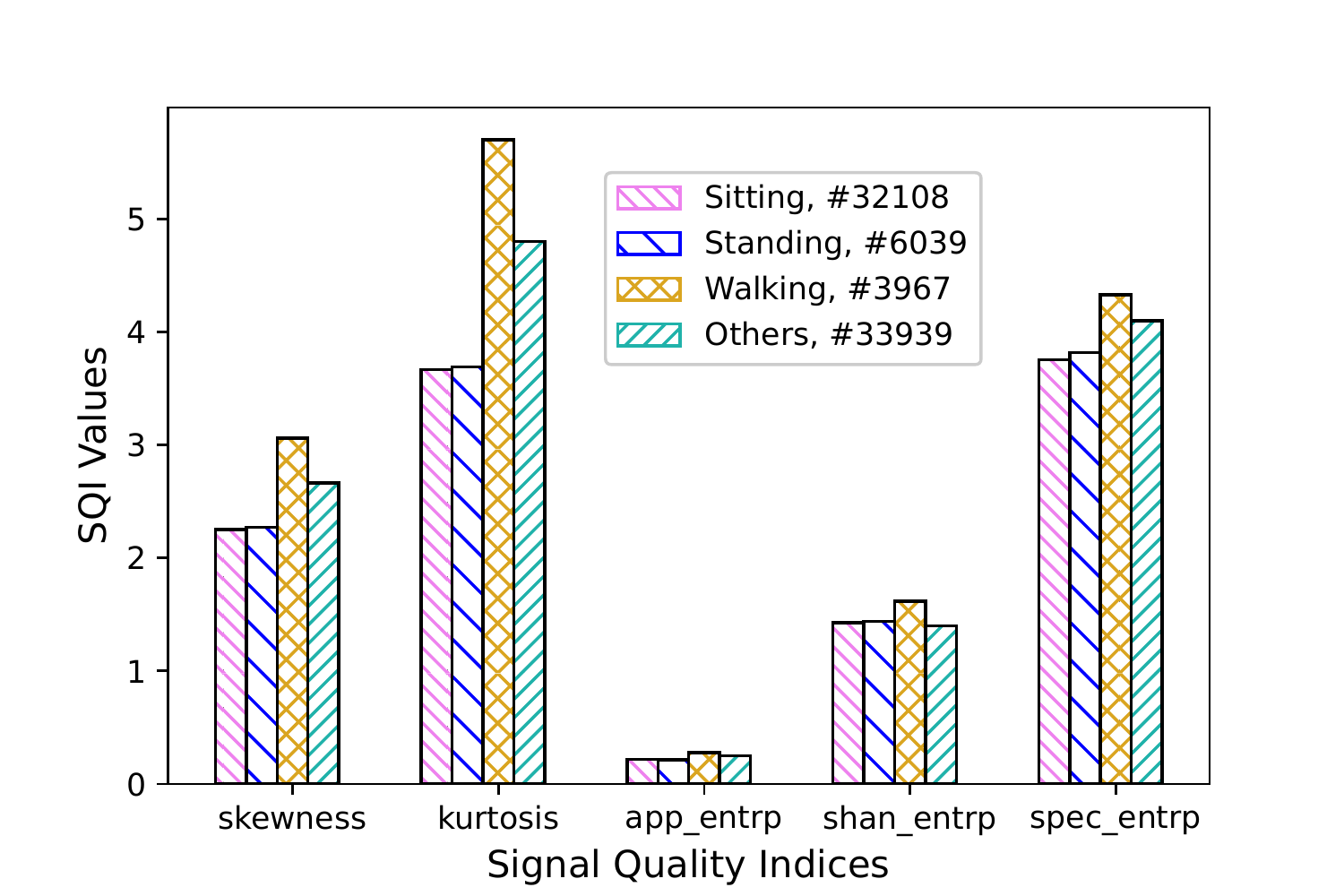}
    \vspace{-1em}
    \caption{Signal Quality Indices for various predicted activities. A lower index means the sample is more reliable.}
    \label{fig: scores}
\end{figure}

The distributions of the five SQI for various activities are shown in Figure~\ref{fig: scores}. 
As expected, with increased motion all the five indices show that data quality decreases among the data marked sitting, standing, walking, respectively. Since the number of samples predicted as "Jogging" was too small, we didn't include that in this analysis. We used the data from all the fourteen subjects.

\subsection{Response Rate and Response time}
\begin{figure}[tbp]
    \centering
    \vspace{-1.5em}
    \includegraphics[width= \msize \linewidth]{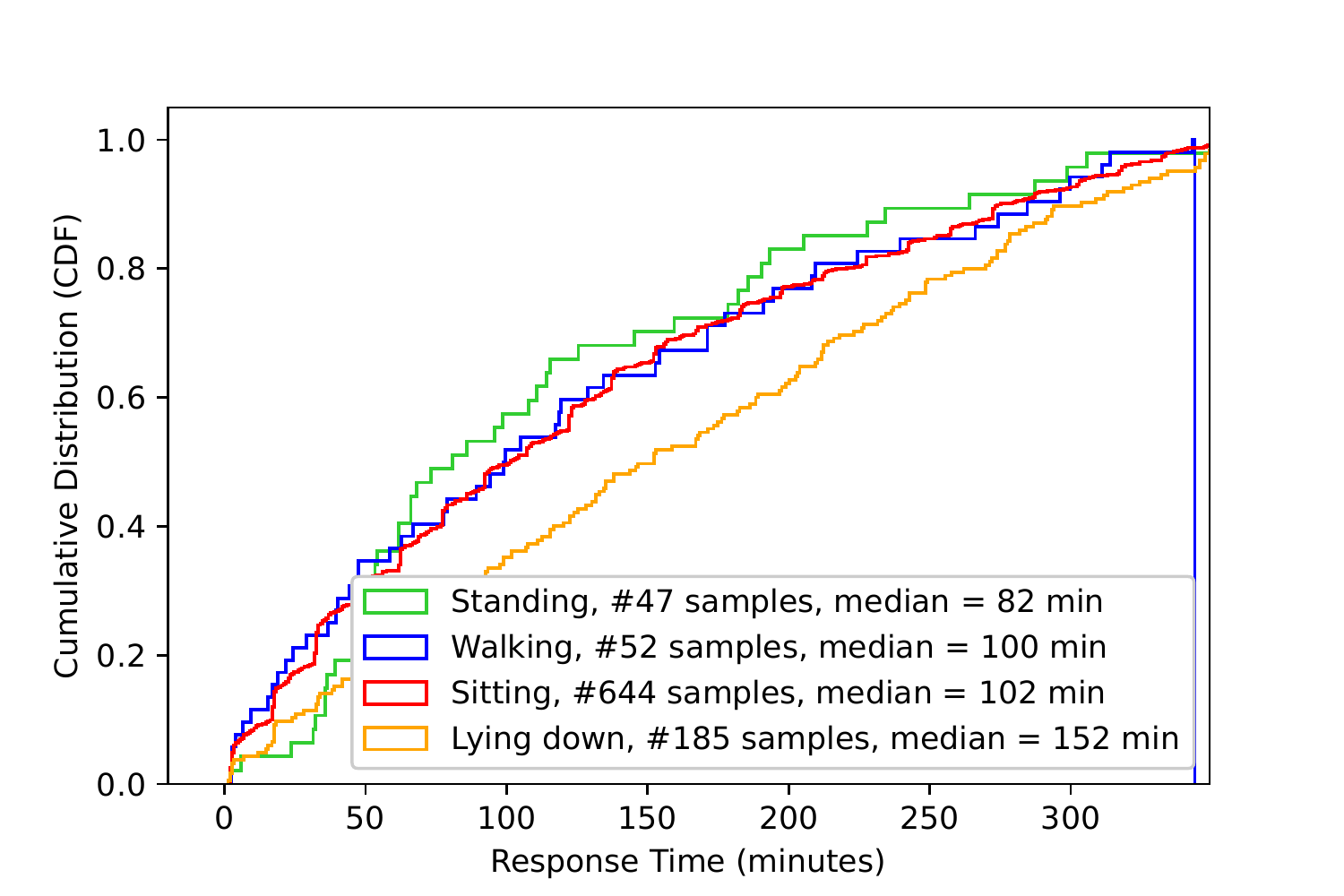}
    \vspace{-1em}
    \caption{Probability of getting a response within a time frame while doing various self reported activities.}
    \vspace{-1.5em}
    \label{fig: response_time}
\end{figure}

Response time is defined as the time it takes a user to respond an EMA after they receive the notification. In Figure \ref{fig: response_time} we present the cumulative density of response time while being in certain contexts (e.g. sitting, etc.). We can see from these CDF plots that users are less likely to respond to the EMA faster when they are \textit{Lying down}, compared to other physical states (or activities).

\begin{figure}[tbp]
    \centering
    %\vspace{-1.5em}
    \includegraphics[width= \msize \linewidth]{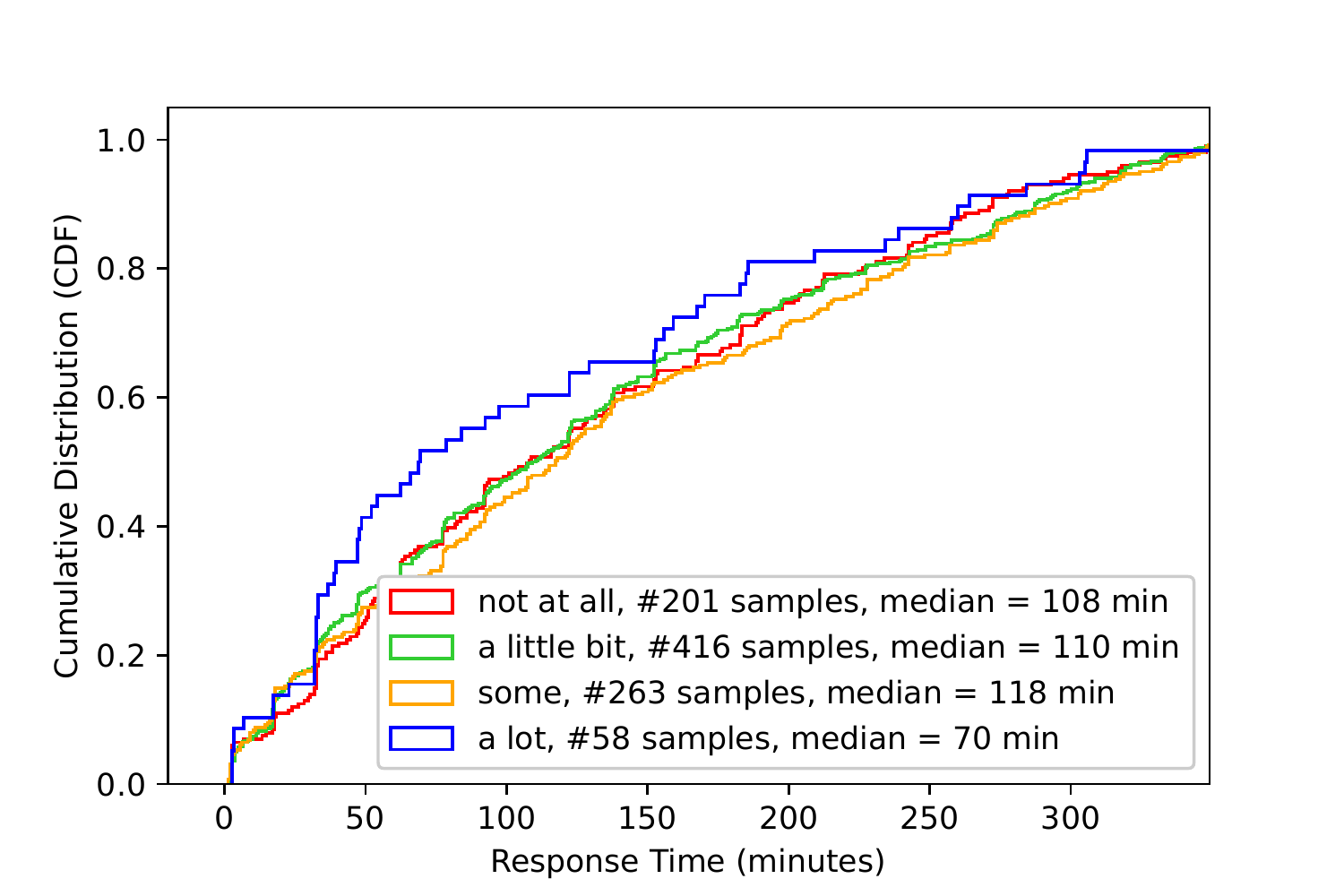}
    \vspace{-1em}
    \caption{Probability of getting a response within a time-frame while in various self reported mental states.}
    \vspace{-1.5em}
    \label{fig: response_time1}
\end{figure}

Similar to the different activities, response time shows different patterns when users are in different reported stress levels. Figure \ref{fig: response_time1} presents the CDF and the median (where the CDF is 0.5) for different self reported stress levels. The pattern suggests that users are more likely to respond faster when they are in stressful situations. 
%and \ref{fig: response_time1}.
% \begin{figure}[tbp]
%     \centering
%     \includegraphics[width=1\linewidth]{Figures/response time over detected stress.pdf}
%     \caption{CDF of response time for detected stress (binary)}
%     \label{fig: response_time2}
% \end{figure}

\begin{figure}[tbp]
    \centering
    \vspace{-1.5em}
    \includegraphics[width= \msize \linewidth]{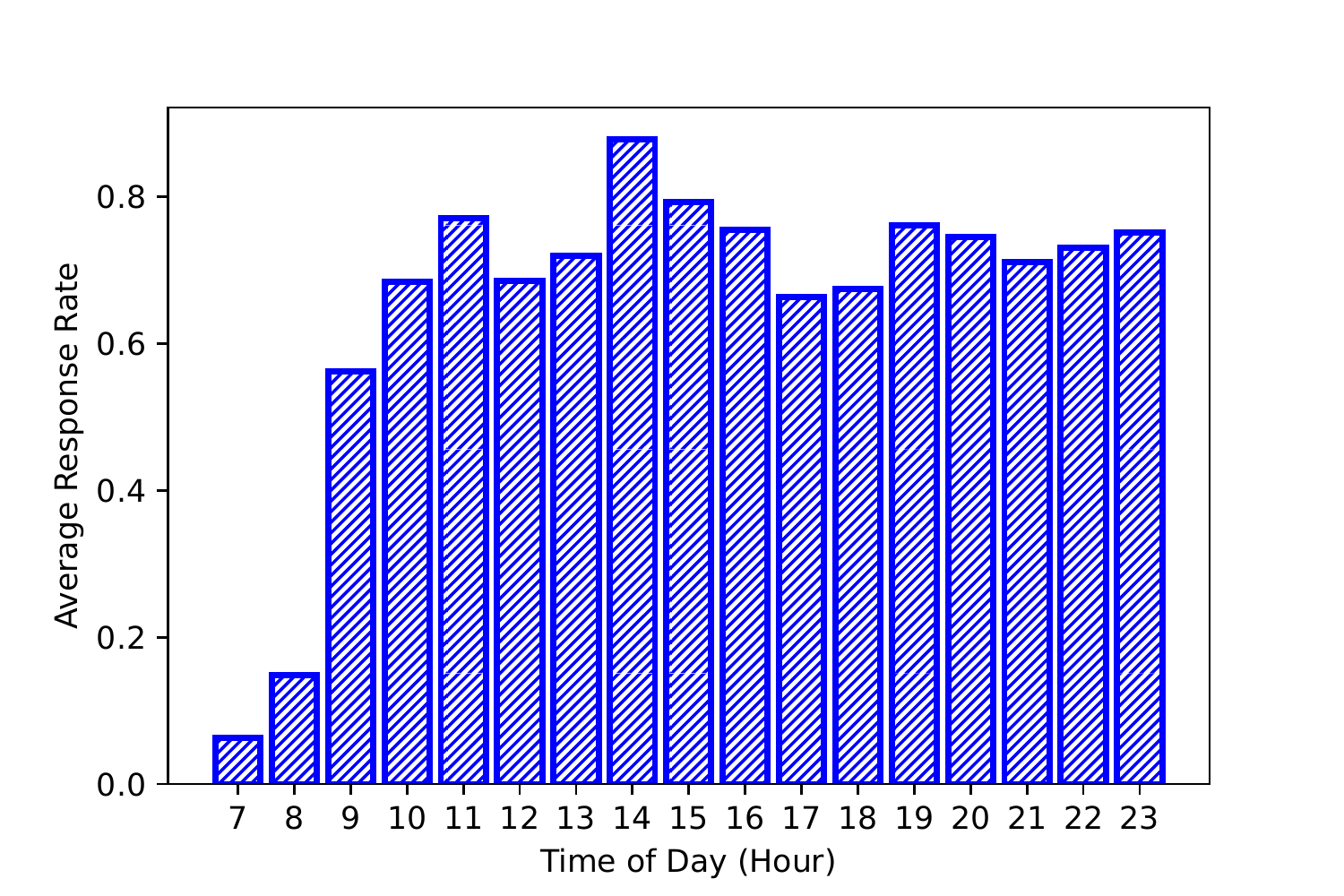}
    \vspace{-1em}
    \caption{Average response rate vs. time of day}
    \vspace{-1em}
    \label{fig: response_rate}
\end{figure}

Response rate, defined as the ratio of the number of responses to the number of queries over a period of time, is also analyzed here. Response rate varies at different hours of day, and also follows different patterns among different subjects, but still some common patterns can be observed among all subjects. Figure \ref{fig: response_rate} shows the average response rate over all the subjects for different hours of day; response rate tends to be lower in early morning and higher in early afternoon. 

\section{Conclusions and Future Work}
\label{sec: conclusion}
Collecting photoplethysmogram (PPG) signals with corresponding self reported labels in everyday settings is a big challenge. 
Our study used the Samsung Gear Sport smart-watch as a sensor device in a system for this phase of data collection and utilized a method to improve the label collection procedure.
The data were collected from fourteen active volunteers in everyday settings. 
We tested our personalized label query engine and performed a set of contextual analysis on the data.
The quality analysis on the biosignals confirms that these signals are more reliable in certain predictable contexts (i.e. in lower physical movements).
A temporal analysis of the data shows that consecutive samples for one subject are similar, but there is no consistent similarity between samples that are more than 15 minutes apart.
In addition, an analysis on users behaviour shows some clear patterns in response times and response rates at different hours of day, and under different mental and physical status.\looseness=-1

These observations motivate our future work that will utilize  more sophisticated methods (possibly variants of active learning) in the labeling process. 
Better labels will allow us to design a classifier that can possibly detect mental health conditions of the users based on biosignals and contextual information -- promising to provide valuable tools for mental health professionals to better diagnose and treat emotional and stress problems in a context-aware and personalized manner.

\begin{acks}
This work was partially supported by NSF Smart and Connected Communities (S\&CC) grant CNS-1831918. 
\end{acks}

\bibliographystyle{ACM-Reference-Format}
\bibliography{main}
\end{document}